\documentclass[journal,twocolumn,howpacs,amsmath,amssymb]{IEEEtran}
\usepackage[dvips]{graphicx} 
\usepackage{amsfonts}
\usepackage{amssymb}
\usepackage{amscd}
\usepackage{amsmath}    
\usepackage{enumerate}
\usepackage{epsfig}
\usepackage{subfigure}
\usepackage{xcolor}
\usepackage{amsthm}
\usepackage{physics}
\usepackage{epstopdf}

\newtheorem{theorem}{Theorem}

\newcommand{\M}{\mathcal{M}}
\newcommand{\upcite}[1]{\textsuperscript{\textsuperscript{\cite{#1}}}}

\begin{document}
\title{{Quantum Coherence and Intrinsic Randomness}}

\author{Xiao Yuan, Qi Zhao, Davide Girolami, Xiongfeng Ma*
\thanks{~X.~Y., Q.~Z., X.~M., Center for Quantum Information, Institute for Interdisciplinary Information Sciences, Tsinghua University, Beijing 100084, China}
\thanks{D.~G., Clarendon Laboratory, Department of Physics, University of Oxford, Parks Road, Oxford OX1 3PU, United Kingdom}
\thanks{*xma@tsinghua.edu.cn}
}
\maketitle

\begin{abstract}
The peculiar uncertainty or randomness of quantum measurements stems from coherence, whose information-theoretic characterization is currently under investigation. Under the resource theory of coherence, it is interesting to investigate interpretations of coherence measures and the interplay with other quantum properties, such as quantum correlations and intrinsic randomness. Coherence can be viewed as the resource for the intrinsic randomness in the measurement outcomes of a state in the computational basis. We observed in our previous work that the coherence of formation, which measures the asymptotic coherence dilution rate, indeed quantifies the uncertainty of a (classical) correlated party about the system measurement outcome. In this work, we re-derive the result from a quantum point of view and then connect the intrinsic randomness to the relative entropy of coherence, another important coherence measure that quantifies the asymptotic distillable coherence. Even though there does not exist bound coherent states, these two intrinsic randomness quantified by coherence of formation and the relative entropy of coherence are different. Interestingly, we show that this gap is equal to the quantum discord, a general form of quantum correlations, in the state of the system of interest and the correlated party, after a local measurement on the former system.
\end{abstract}


\section{Introduction}
According to the Born rule\upcite{born}, the outcome of a quantum measurement is intrinsically random. Given a quantum state $\ket{\alpha}=\sum_i c_i \ket i$, where  $\{c_i\}$ are complex coefficients, the result of a projection measurement $\{\dyad{i}\}$ is not deterministic, having the output $i$ with a probability $p_i=|c_i|^2$. Such randomness differs from the classical uncertainty due to uncharacterized measurements.
This intrinsic randomness promises to be a potential resource for information processing tasks. There are many proposals for quantum random number generation, we refer to Refs.\upcite{Ma2016QRNG,RevModPhys.89.015004} for reviews. As it is immediately clear from the example above, intrinsic randomness is a consequence of breaking coherent superpositions of quantum states, a phenomenon nowadays routinely observed in the laboratory. Recently, several works have studied the properties of coherent states as an information-theoretic resource\upcite{spekkens,aberg}. It turns out that the key notion to identify a resource, the definition of free operations, is not unique in the case of coherence. As a result, several measures have been proposed to quantify coherent superposition.

The most intuitive way to quantify coherence is via the distance to the set of incoherent states ${\cal I}$ for a reference basis $\{\ket{i}\}$, given by an appropriate yet arbitrary (pseudo-)metric function,
\begin{equation} \label{eq:geomeasure}
\begin{aligned}
C_d(\rho)=\min\limits_{\sigma\in {\cal I}} d(\rho,\sigma),
\end{aligned}
\end{equation}
where $d(\rho,\sigma)$ is a function to measure the distance of two states and $\cal I$ is the incoherent state set which contains all statistical mixtures of the basis states $\{\dyad{i}\}$. We label this notion of coherence as the BCP coherence\upcite{Baumgratz14,herbut}. A widely employed solution is to adopt the relative entropy of coherence as a measure,
\begin{equation} \label{eq:RelEntropy}
\begin{aligned}
C_R(\rho)=\min\limits_{\sigma\in {\cal I}} S(\rho||\sigma),
\end{aligned}
\end{equation}
where the relative entropy of two states are given by $S(\rho||\sigma)=\tr(\rho\log (\rho)-\rho\log(\sigma))$, mainly because of its computability and importance in information theory\upcite{vedral}.
Another option is to quantify coherence via a convex-roof construction, called the coherence of formation, via
\begin{equation}
	C_{f}(\rho) \equiv \min _{\left\{p_{j}, | \psi_{j}\right\rangle \}} \sum_{j} p_{j} C_R\left(\dyad{\psi_{j}} \right),
\end{equation}
where the minimisation is over all possible decomposition of $\rho=\sum_j p_j\dyad{\psi_{j}}$.
In the resource theory of coherence, the relative entropy of coherence and the  coherence of formation measures the asymptotic coherence distillation and dilution rates, respectively\upcite{winter2016}.
The coherence distillation and dilution problems are then extended into the non-asymptotical scenario using other coherence measures\upcite{Zhao2018OneShot,PhysRevLett.121.010401,zhao2019one,liu2018quantum}. We refer to Ref.\upcite{RevModPhys.89.041003,hu2018quantum} for reviews of the resource theory of coherence.
While the parent notion of asymmetry has a clear-cut interpretation in a number of physical settings\upcite{spekkens} and other significant advances have been reported\upcite{yadin}, the operational power offered by the BCP coherence still needs to be fully understood.

Given an input quantum state and a measurement owned by Alice, the intrinsic randomness of outcome, against a potential adversary Eve, is a topic of interest in the quantum information theory. From Alice's point of view, in the asymptotic limit, the Shannon entropy $H(p_i)_{\rho}=-\sum_i p_i \log p_i$ is the quantifier of the total uncertainty of a measurement with probability distribution $\{p_i\}$ in the measurement outcome of a state $\rho$, named \emph{nominal randomness}. The nominal randomness consists of two parts, \emph{intrinsic randomness} and \emph{extrinsic randomness}. The intrinsic randomness is quantum randomness which should be unpredictable, while the extrinsic randomness can be predicated by a quantum correlated party Eve in principle. For example, for pure states, since Eve's system is decoupled from Alice's one, the only kind of uncertainty is the truly quantum one (intrinsic randomness), there is no extrinsic randomness in the measurement outputs. Thus, the Shannon entropy (nominal randomness) is itself a measure of intrinsic randomness. For the case of incoherent states, the measurement uncertainty is purely classical (extrinsic randomness) and there is no intrinsic randomness as it entirely depends on Alice's incomplete knowledge of her system. Our goal is to quantify the quantum intrinsic randomness concerning the existence of a potential adversary by coherence measures and explore the quantum contribution to the total uncertainty. We consider a scenario which is consistent with the aforementioned two situations and allows to give an operational interpretation to the quantum coherence in the more complex case of mixed coherent states. From an operational perspective, the extractable randomness is measured by the conditional min-entropy\upcite{konig2009operational}. We therefore also show how to obtain our results by considering the asymptotic limit with the conditional min-entropy.

In this work, we focus on the interplay between quantum coherence and intrinsic randomness. In particular, we study operational interpretations of the relative entropy of coherence and the coherence of formation in characterizing intrinsic randomness. The result for the coherence of formation has been previously considered\upcite{Yuan2015}. This work re-derives this result by focusing on a more rigorous scenario with the conditional min-entropy. We further consider a more general scenario and relate the relative entropy of coherence with intrinsic randomness. We also found that while there is no bound coherent states, which have non vanishing coherence of formation but zero coherence of distillation, the two quantities are different.

 The strategy we adopt is presented as follows. In Section~\ref{qrand}, we consider a projection measurement of the quantum state in the reference basis. We pick the smooth conditional min-entropy as the quantifier of the total uncertainty of the measurement outcomes conditioned on all possible environment systems. To do so, we consider a bipartite extension of a system manipulated by Alice, say accessible to a pair Alice-Eve in the state $\rho_{AE}$, and address the question of how much information Eve can access about Alice's measurement outcome with the probability distribution $\{p_i\}$ and outputs being the elements of a reference basis $\{\ket{i}_A\}$. We show that in the asymptotic scenario, Eve's ignorance is quantified by the relative entropy of coherence of Alice's state with respect to the reference basis, which is a good quantifier of the quantum uncertainty on Alice's measurement: $\min\limits_{\rho_E}H(\{p_i\}|E)_{\rho_{AE}}=C_R(\rho_A)$.

We then compare the results with the scenario where Eve gains information about Alice's measurement  ``classically'' by performing a measurement on her part (Section \ref{crand}). A previous work proved that, as in the former setting, Eve's uncertainty on Alice's outcome is a full-fledged measure of the BCP coherence, namely, the coherence of formation\upcite{Yuan2015, liu2018superadditivity}.  Such a  measure is  obtained by a convex roof construction, which is different from the relative entropy of coherence.
In this work, we re-derive the same result with the smooth conditional entropy.
Furthermore, we show that the gap between the two quantities, which characterizes the irreversibility of coherence resource theory\upcite{winter2016}, corresponds to the quantum discord of the Alice-Eve's system {after Alice's measurement} (Section \ref{disc}). This is an interesting result as the state is separable (precisely, it is a classical-quantum state) so no entanglement appears and the quantum advantage of Eve is indisputably due to quantum discord. In Section \ref{concl}, we draw our conclusions.

\section{Coherence and intrinsic randomness}
In this section, we introduce the intrinsic randomness or uncertainty that one has conditioned on a correlated party. We show that the intrinsic randomness is quantified by the coherence of the state in the measurement basis.

\subsection{Relative entropy of coherence as uncertainty of correlated party}\label{qrand}
Let us consider a $d$-dimensional Hilbert space and a reference  basis  $I:=\{\ket{i}\} = \left\{\ket{1},\ket{2},\dots,\ket{d}\right\}$. Suppose a projective measurement $\{\dyad{i}\}$ is performed on a given quantum state $\rho_A$ accessed by Alice. The measurement outcome has a probability distribution $\{p_i\}$, with $\sum_{i=1}^d p_i=1$ and $p_i= \textrm{Tr}[\rho \dyad{i}]\geq 0 $. We aim to assess the intrinsic or unpredictable randomness of the measurement outcome. To do so, we consider another adversarial party Eve where the joint state shared by Alice and Eve is $\rho_{AE}$, satisfying $\tr_E[\rho_{AE}]=\rho_A$ with partial trace over system $E$. Note that the state $\rho_{AE}$ is not assumed to be pure in our analysis, though we show shortly that considering pure states $\rho_{AE}$ is sufficient for characterising intrinsic randomness. The measurement can be represented as a dephasing channel
\begin{equation}
\Delta_A(\rho) = \sum_i\bra{i}\rho\ket{i}\dyad{i},
\end{equation}
and the joint state  after Alice's measurement becomes $\rho'_{AE}=\Delta_A(\rho_{AE})$. The state $\rho'_{AE}$ is a classical-quantum state and the randomness of the measurement outcome conditioned on Eve's system is characterized by the smooth conditional min-entropy\upcite{konig2009operational},
\begin{equation} \label{eq:smoothmin}
H_{\min}^\varepsilon(A|E)_{\rho'_{AE}} = \textrm{sup}_{\|\sigma_{AE}-\rho'_{AE}\|\le\varepsilon}H_{\min}(A|E)_{\sigma_{AE}},
\end{equation}
where  $\varepsilon$ is the smooth parameter, and the supremum takes over all states $\sigma_{AE}$ that are $\varepsilon$ close to $\rho'_{AE}$ with $\|\sigma_{AE}-\rho'_{AE}\|=1-F(\sigma_{AE},\rho'_{AE})$ and fidelity $F(\rho,\sigma)=(\tr[\sqrt{\sqrt{\rho}\sigma\sqrt{\rho}}])^2$. Here the conditional min-entropy $H_{\min}(A|E)_{\rho_{AE}}$ is
\begin{equation}
	H_{\min}(A|E)_{\rho_{AE}} = -\textrm{inf}_{\sigma_E}D_{\max}(\rho_{AE}\|\textrm{id}_A\otimes \sigma_E),
\end{equation}
where the infimum is over all normalized density operators on system $E$, $\textrm{id}_A$ is the identity matrix on system $A$, and the max-relative entropy $D_{\max}(\rho\|\sigma)$ is defined by
\begin{equation}
	D_{\max}(\rho\|\sigma) = \textrm{inf}\{\lambda\in \mathbb R, \rho\le2^\lambda\sigma\}.
\end{equation}
We choose the smooth conditional min-entropy which measures the maximum amount of private and uniformly random bits that can be extracted\upcite{konig2009operational}. Specifically, given a general classical-quantum state $\rho_{AE}$, one can apply an extractor on system $A$ so that it is $\varepsilon'$-close to the perfectly uniform bits that are independent of any side information of system $E$. Therefore, the length $\ell_{\mathrm{extr}}^{\varepsilon'}(A | E)$ of the extracted bits is given by,
\begin{equation}
\ell_{\mathrm{extr}}^{\varepsilon'}(A | E)=H_{\min }^{\varepsilon}(A | E)_{\rho_{AE}}+O(\log 1 / \varepsilon'),
\end{equation}
with $\varepsilon \in\left[\frac{1}{2} \varepsilon^{\prime}, 2 \varepsilon^{\prime}\right]$. One can also consider other entropic quantifiers which may characterise other operational tasks. We leave the discussion of generalising our results to other entropic quantifiers to a future work.


\begin{figure}[t]
\centering
{\includegraphics[width=0.35\linewidth]{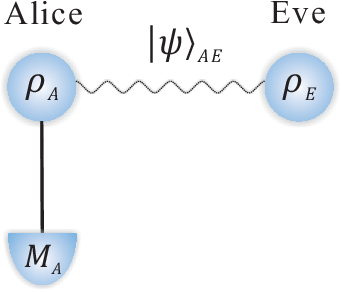}}\quad\quad
{\includegraphics[width=0.35\linewidth]{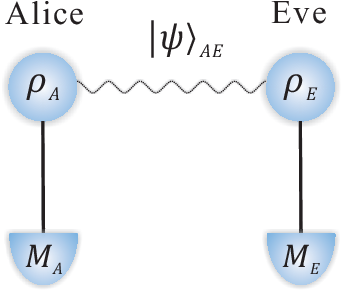}}\\
\quad(a)\quad\quad\quad\quad\quad\quad\quad\quad\quad(b)
\caption{Intrinsic randomness of measurements conditioned on quantum and classical information.
(a)
In a bipartite Alice-Eve system described by a pure state $\psi_{AE}$, the quantum uncertainty of a measurement performed by Alice on the system in the mixed state $\rho_A$ is given by the amount of uncertainty Eve has on the measurement outcome. Such quantum uncertainty is quantified by the relative entropy of coherence $R_I^Q(\rho_A)$.
(b)
Alternative definition of quantum coherence as uncertainty.  In a bipartite Alice-Eve system described by a pure state $\psi_{AE}$, the quantum uncertainty of a measurement performed by Alice on the system in the mixed state $\rho_A$ is given by the minimum amount of uncertainty Eve has on the measurement outcome {\it after performing a measurement on her own systems}. Such quantum uncertainty is quantified by the convex roof measure $R^C_I(\rho_A)$.}\label{qfigure}
\end{figure}

After Alice's measurement on $\rho_{AE}$, the randomness of the measurement can be characterized by
\begin{equation}
	R_I^{Q,\varepsilon}(\rho) = \min_{\rho_{AE}:\tr_E[\rho_{AE}]=\rho_A} H_{\min}^\varepsilon(A|E)_{\Delta_A(\rho_{AE})},
\end{equation}
where the minimization is over all states $\rho_{AE}$ satisfying $\tr_E[\rho_{AE}]=\rho_A$.
For each $\rho_{AE}$, we can further consider its purification by introducing an additional system $E'$ so that the whole system is $\ket{\psi}_{AEE'}$, satisfying $\tr_{E'}[\dyad{\psi}_{AEE'}]=\rho_{AE}$. Then the intrinsic randomness becomes
\begin{equation}
\begin{aligned}
		R_I^{Q,\varepsilon}(\rho)&= \min_{\ket{\psi}_{AEE'}:\tr_{EE'}[\dyad{\psi}_{AEE'}]=\rho_{A}}H_{\min}^\varepsilon(A|E)_{\Delta_A\circ\tr_{E'}(\ket{\psi}_{AEE'})},\\
		&\ge \min_{\ket{\psi}_{AE}:\tr_{E}[\dyad{\psi}_{AE}]=\rho_{A}}H_{\min}^\varepsilon(A|E)_{\Delta_A(\ket{\psi}_{AE})},
\end{aligned}
\end{equation}
where the second inequality is because of the data-processing inequality of the smooth conditional min-entropy defined in Eq.~\eqref{eq:smoothmin}. As the minimisation of the second line is only a special case of the minimisation in the definition of $R_I^Q$ in Eq.~\eqref{Eq:defiRIQ}, we have
\begin{equation}
	R_I^{Q,\varepsilon}(\rho) = \min_{\ket{\psi}_{AE}:\tr_{E}[\dyad{\psi}_{AE}]=\rho_{A}}H_{\min}^\varepsilon(A|E)_{\Delta_A(\ket{\psi}_{AE})}.
\end{equation}
As all purification states $\ket{\psi}_{AE}$ are equivalent under isometry on system $E$, which nevertheless does not affect the smooth conditional min-entropy, we therefore have
\begin{equation}
	R_I^{Q,\varepsilon}(\rho) = H_{\min}^\varepsilon(A|E)_{\Delta_A(\ket{\psi}_{AE})},
\end{equation}
where $\ket{\psi}_{AE}$ is any purification of $\rho_A$ as shown in Fig.~\ref{qfigure}(a).

Suppose Alice prepares $n\gg1$ copies of $\rho_A$ and performs the projective measurement for all the copies, the average randomness of each measurement outcome with the limit $n\rightarrow\infty$ and $\varepsilon\rightarrow0^+$ is then characterized by
\begin{equation}
\begin{aligned}
		R_I^Q &= \lim_{\varepsilon\rightarrow0^+}\lim_{n\rightarrow\infty}\frac{1}{n}R_I^{Q,\varepsilon}(\rho^{\otimes n}),\\
		&=\lim_{\varepsilon\rightarrow0^+}\lim_{n\rightarrow\infty}\frac{1}{n}H_{\min}^\varepsilon(A^{n}|E^n)_{\Delta^{\otimes n}_A(\ket{\psi}^{\otimes n}_{AE})}.
\end{aligned}
\end{equation}

\begin{theorem}
The intrinsic randomness of Alice's measurement outcome conditioned on any (quantum) adversary Eve, $R^Q_I $ is quantified by the relative entropy of coherence,
\begin{equation} \label{Eq:rq}
\begin{aligned}
R^Q_I(\rho_A) &=C_R(\rho_A)\\
&=S(\rho_A^{\mathrm{diag}}) - S(\rho_A) \\
&=S(\rho_A||\rho_A^{\mathrm{diag}}). \\
\end{aligned}
\end{equation}
\end{theorem}

\begin{proof}
According to the asymptotic equipartition property of the smooth entropies\upcite{tomamichel2015quantum}, we have
\begin{equation}\label{Eq:AEP}
	\lim_{\varepsilon\rightarrow0^+}\lim _{n \rightarrow \infty}\left\{\frac{1}{n} H_{\min }^{\varepsilon}\left(A^{n} | B^{n}\right)_{\rho^{\otimes n}}\right\} = H(A | B)_{\rho}.
\end{equation}
Therefore, we have
\begin{equation}\label{Eq:defiRIQ}
	R_I^Q = H(A|E)_{\Delta_A(\ket{\psi}_{AE})},
\end{equation}
where $H(A|B)_{\rho_{AB}}=S(\rho_{AB})-S(\rho_{B})$ is the von Neumann conditional entropy, $S(\rho)=-\tr[\rho\log\rho]$ is the von Neumann entropy and $\rho_B=\tr_A[\rho_{AB}]$. The right hand side of Eq.~\eqref{Eq:defiRIQ} can be explicitly evaluated for example with the analysis in Ref.\upcite{Coles12}. We also briefly summarize the proof here for self-consistence. After the measurement, the state is changed to
\begin{equation}
	 \rho_{AE}' = \Delta_A(\ket{\psi}_{AE}) = \sum_i p_i\ket{i}_A\bra{i}\otimes\rho_i^E,
\end{equation}
where  $\rho_E  = \sum_i p_i\rho_i^E$ and each $\rho_i^E=\bra{i}_A(\dyad{\psi}_{AE})\ket{i}_A/p_i$ is a pure state.  Using the  equality $S\left(\sum_ip_i\dyad{i}\otimes\rho_i\right) = H(p_i) + \sum_i p_i S(\rho_i)$, the conditional entropy of the post-measurement state is then
\begin{equation}
	S(A|E)_{\rho_{AE}'} = H(p_i) + \sum_ip_iS(\rho_i^E)  - S(\rho_E).
\end{equation}
Since $H(p_i) = S(\rho_A^{\mathrm{diag}})$ with $\rho_A^{\mathrm{diag}}:=\sum_i p_i \dyad{i}$, $S(\rho_E) = S(\rho_A)$, and $S(\rho_i^E) = 0, \forall i,$ we obtain  our result in the theorem.

\end{proof}
Therefore, as a measure of BCP coherence\upcite{Baumgratz14}, the relative entropy of coherence $C_R(\rho_A)$ satisfies all the requirements for a consistent measure of intrinsic randomness.

\subsection{Coherence of formation as uncertainty of correlated party}\label{crand}
We observed that the quantum uncertainty of a local measurement corresponds to the best case uncertainty of a correlated party Eve, as quantified by means of  the quantum conditional entropy.  We compare the result with an alternative measure of quantum coherence --- coherence of formation\upcite{Yuan2015}. The setting is for the sake of clarity depicted in Fig.~\ref{qfigure}(b). The difference is that Eve performs a measurement whose outcomes follow a probability distribution $\{q^E_i\}, q^E_i=\textrm{Tr}[\rho_E\ket{e_i'}_E\bra{e_i'}]$, on her own system to classically predict Alice's measurement outcome.
Suppose Eve's measurement is represented by a quantum channel as
\begin{equation}
	\M(\rho) = \sum_i \tr[\rho O_i]\dyad{i},
\end{equation}
where $O_i\ge 0$, $\sum_i O_i = \textrm{id}_E$, and $\textrm{id}_E$ is the identity matrix of system $E$. For one copy of $\rho_{AE}$, the randomness of Alice's measurement outcome conditioned on Eve's measurement outcome is then
\begin{equation}
	R_I^{C,\varepsilon}(\rho) = \min_{\M}\min_{\rho_{AE}:\tr_E[\rho_{AE}]=\rho_A} H_{\min}^\varepsilon(A|E)_{\Delta_A\otimes \M(\rho_{AE})},
\end{equation}
where the minimisation is also over all Eve's possible measurements and all possible $\rho_{AE}$ satisfying $\tr_E[\rho_{AE}]=\rho_A$. With a similar argument of the proof in the last section, we only need to focus on any one specific purification $\ket{\psi}_{AE}$ of $\rho_A$. Therefore, the intrinsic randomness is
\begin{equation}
	R_I^{C,\varepsilon} (\rho)= \min_{\M} H_{\min}^\varepsilon(A|E)_{\Delta_A\otimes \M(\ket{\psi}_{AE})}.
\end{equation}

When Alice prepares $n\gg1$ copies of $\rho_A$ and performs the projective measurement for all the copies, the average randomness of each measurement outcome is
\begin{equation}
	R_I^C = \lim_{\varepsilon\rightarrow0^+}\lim_{n\rightarrow\infty}\frac{1}{n}R_I^{C,\varepsilon}(\rho^{\otimes n}).
\end{equation}
In general, Eve's measurement can be a joint measurement on all her local systems. In this case, we have $R_I^C=R_I^Q$. Here instead, we restrict to the scenario that Eve also performs identical measurement for each copy of her local system\footnote{Note that here we only need to assume that Eve performs independent measurements on each copy of her local system. It reduces to the identical measurement case by considering a random permutation of all the states.}. Therefore the joint state after the measurements is ${\Delta_A^{\otimes n}\otimes \M^{\otimes n}(\ket{\psi}_{AE}^{\otimes n})}$.
According to the asymptotic equipartition property of the smooth entropies\upcite{tomamichel2015quantum} shown in Eq.~\eqref{Eq:AEP}, we have
\begin{equation}\label{Eq:defiRIC}
	R_I^C = \min_{\M}H(A|E)_{\Delta_A\otimes \M(\ket{\psi}_{AE})}.
\end{equation}
According to our recent work\upcite{liu2018quantum}, we can conclude as follows.
\begin{theorem}
The intrinsic randomness of Alice's measurement outcome conditioned on any (classical) adversary Eve with independent measurements is quantified by the coherence of formation,
\begin{equation}
	R_I^C = C_{f}(\rho) \equiv \min _{\left\{p_{j}, | \psi_{j}\right\rangle \}} \sum_{j} p_{j} S\left(\Delta\left( \dyad{\psi_{j}} \right)\right),
\end{equation}
where $C_{f}(\rho)$ is the coherence of formation and the minimisation is over all decomposition of $\rho=\sum_{j} p_{j} \dyad{\psi_{j}}$.
\end{theorem}

\subsection{Qubit calculation for $R_I^Q$ and $R_I^C$}
\begin{figure}[t]
\centering
\resizebox{8cm}{!}{\includegraphics[scale=1]{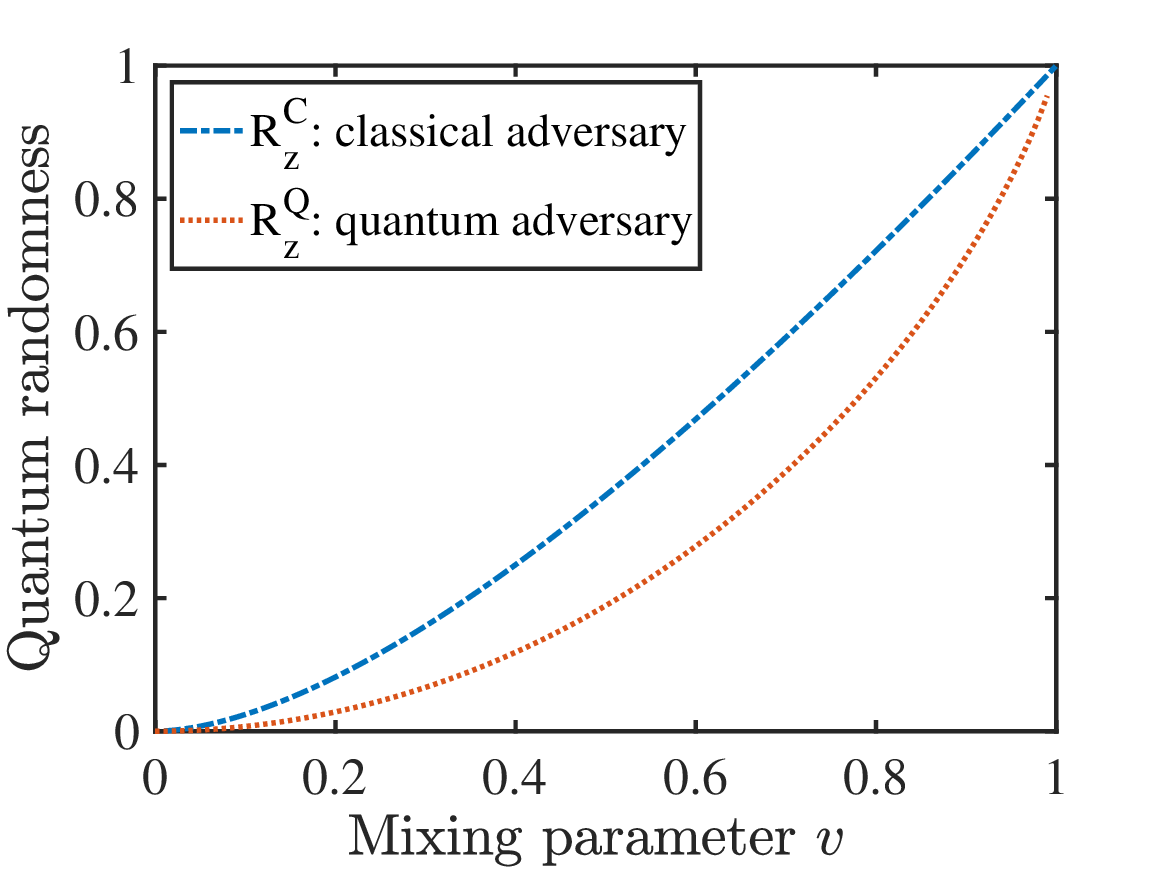}}
\caption{Comparison of the measures of quantum uncertainty $R_z^Q$ (red dotted line) and $R_z^C$ (blue dot-dashed line) in the qubit state $\rho_A(v)=v\ket{+}\bra{+} + \frac{1 - v}{2}\mathbb{I}$ versus the mixing parameter $v$. $R_z^Q$ is the intrinsic randomness of Alice's measurement outcome conditioned on a quantum adversary; $R_z^C$ is the intrinsic randomness conditioned on a classical adversary.}\label{fig:RQC}
\end{figure}

The quantum uncertainty measure obtained by convex roof extension   is a measure of BCP coherence as well\upcite{Yuan2015}. Let us compare  the two quantities $R^C_I(\rho_A)$ and  $R^Q_I(\rho_A)$ in a simple example about  a qubit system.  In the Bloch sphere representation, $\rho_A = (I+\vec{n}\cdot\vec{\sigma})/2$, where $\vec{n} = (n_x,n_y,n_z)$ and $\vec{\sigma} = (\sigma_x,\sigma_y, \sigma_z)$ are the Pauli matrices. Supposing that the measurement basis is the $\sigma_z$ eigenbasis, which is denoted by $\{\ket{0},\ket{1}\}$, then we obtain
\begin{eqnarray}\label{Eq:ef}
R_{z}^C(\rho_A)&=& H\left(\frac{1+\sqrt{1-n_x^2 - n_y^2}}{2}\right)\\
R_{z}^Q(\rho_A)&=& H\left(\frac{n_z + 1}{2}\right) - H\left(\frac{|n| + 1}{2}\right),\nonumber
\end{eqnarray}
where $|n| = \sqrt{n_x^2+n_y^2+n_z^2}$ and $H$ is the binary entropy.
Specifically, for the state $\rho_A(v) = v\dyad{+} + \frac{1 - v}{2}I$, where $\ket{+} = (\ket{0}+\ket{1})/2, v\in[0,1], \vec{n}(v) = (v, 0, 0)$, we have
\begin{eqnarray}\label{Eq:ef}
R_{z}^C(\rho_A)&=& H\left(\frac{1+\sqrt{1-v^2}}{2}\right),\\
R_{z}^Q(\rho_A)&=& 1 - H\left(\frac{v + 1}{2}\right).
\end{eqnarray}
In Fig.~\ref{fig:RQC}, we plot the two measures versus the mixing parameter $v$. By definition, the randomness quantifier $R_z^C$ is against a classical adversary, who can only perform independent measurements on her local systems. On the contrary, the randomness quantifier $R_z^Q$ assumes a powerful quantum adversary, who can perform general measurements. As a classical adversary is a special case of a general quantum adversary, the quantum coherence measure $R_z^Q$ is  generally smaller than  $R_z^C$. As they both measure randomness, it is not hard to see that  they both vanish when the state is incoherent and they converge to the Shannon entropy in the pure state case. All those intuitions are verified in the numerical example shown in Fig.~\ref{fig:RQC}.

\section{Quantum Coherence gap and Quantum Discord}


\subsection{Quantum Discord as difference between coherence measures}\label{disc}
Note that the quantum coherence measure $R_I^Q(\rho_A)$ is obtained when considering an adversary that utilizes quantum information to predict Alice's measurement outcome. In comparison, the measure $R_I^C(\rho_A)$ is obtained when the adversary performs independent and identical measurements on her local systems to have a classical prediction. Obviously, the latter adversary strategy is a special case of the former one, and hence $R_I^C(\rho_A) \geq R_I^Q(\rho_A)$.
In general, there is a non-zero gap between the two  quantum coherence measures, while they both vanish for incoherent states. As the difference between the two frameworks in Fig.~\ref{qfigure} is brought about by making a measurement on Eve's party, it is intuitive to think that the gap is related to how much the local measurement changes the state. Indeed, we show that such a gap is associated to the quantum discord of the bipartite state $\rho_{AE}'=\sum_i p_i \dyad{i}\otimes \rho_E^i$ of the system after Alice carried out her measurement.  Discord (we omit the quantum label from now on) is a kind of quantum correlation which equals entanglement for pure states but also shows up in all but a null measure set of separable states. It can be interpreted as the minimum disturbance induced on a bipartite system by a local measurement\upcite{Modi12}, but here it quantifies the advantage of a quantum correlated system Eve in accessing information about Alice's measurement. Its peculiarity is its asymmetry, as a measurement on one party has in general a different effect than performed on a different subsystem.
For a state $\rho_{AE}$, the discord defined as \[D_E(\rho_{AE})=\min\limits_{\{q_i^E\}}S(A|\{q_i^E\})_{\rho_{AE}}-S(A,E)_{\rho_{AE}}+S(E)_{\rho_{AE}}\] measures the least possible disturbance of a measurement with probability distribution $\{q_i^E\}$ on the $E$ party. Simple algebra steps show that $\min\limits_{\{q^E_i\}}S(A|\{q^E_i\})_{\rho_{AE}'}=\min\limits_{\{q^E_i\}}H(\{p_i\}|\{q^E_i\})_{\psi_{AE}}$. Hence, we obtain the following result for the meaning of the gap of the two coherence measures.

\begin{theorem}
The gap between the relative entropy of coherence and the coherence of formation is given by the discord of the joint state after Alice's measurement, i.e. the least possible state change induced by an Eve's  measurement,
\begin{equation}\label{Eq:discord}
  R_{I}^C(\rho_A) - R_{I}^Q(\rho_A)  = D_E(\rho_{AE}').
\end{equation}
\end{theorem}

\begin{table}[t]
\centering
\caption{Comparison between coherence and entanglement  measures. COF: Coherence of formation; REC: Relative entropy of coherence}
\begin{tabular}{ccc}
  \hline
  Properties&Coherence/Uncertainty&Entanglement\\
  \hline
  Cost& COF $R^C_I$, Eq.~\eqref{Eq:defiRIC}&Entanglement of formation\\
  Distillation&REC  $R^Q_I$, Eq.~\eqref{Eq:defiRIQ}& Distillable Entanglement  \\
  Gap&Discord, Eq.~\eqref{Eq:discord} & Bound Entanglement\upcite{jaeger}\\
  \hline
\end{tabular}\label{table}
\end{table}

We observe that, in the resource theory of quantum coherence, the coherence of formation and the relative entropy of coherence measure the coherence cost and the distillable coherence in the asymptotic limit, respectively\upcite{winter2016}. Thus, the coherence cost and the distillable coherence equal the quantum uncertainty conditioned on Eve's classical\upcite{Yuan2015} and  quantum strategies here discussed.  The scenario is similar to what happens in the entanglement resource theory\upcite{Horodecki09}, where there  is a nonzero gap between the entanglement cost and the distillable entanglement (Table \ref{table}). In particular, some entangled states have zero distillable entanglement, a phenomenon called bound entanglement. However, a key difference is that there is no  coherent states with zero coherence of distillation\upcite{winter2016}. Hence, it emerges that zero relative entropy of coherence on a local Alice's measurement  implies zero coherence cost, $R_I^Q(\rho_A)=0\Rightarrow R_I^C(\rho_A)=0$  and then zero quantum discord, i.e. there exists at least a measurement  on Eve's side which does not change the state.  We also observe that the state $\rho_{AE}'$ is always separable. Thus, the quantum advantage in accessing non-local information about a correlated party measurement is here genuinely due to quantum discord, rather than entanglement.

\subsection{An example}
To clarify the result, we consider the following example inspired by the cryptographic scenario of  the BB84  protocol\upcite{bb84}.   Alice processes two bits information representing eigenbasis and polarization of a quantum state $\rho_A$. If the basis bit is $0$ ($1$), she  prepares the state in the $X$ ($Z$) basis, while if the polarization bit is $0$ ($1$), the state has polarization up (down) in the chosen eigenbasis. To set the notation, if the two bits are $00, 01, 10, 11$, Alice prepares $\ket{0},\ket{1},\ket{+},\ket{-}$, respectively. Let us suppose the probability of choosing each state is equal,  and that Alice sends the quantum state to Eve, who tries to guess the state. Then, the state shared by Alice and Eve is given by
\begin{eqnarray}\label{state}
\rho_{AE}'&=&\frac{1}{4}(\dyad{00}\otimes\dyad{0}+\dyad{01}\otimes\dyad{1}\\
&+&\dyad{10}\otimes\dyad{+}+\dyad{11}\otimes\dyad{-}).\nonumber
\end{eqnarray}
Equivalently, we can consider that Alice and Eve initially share a pure state
\begin{equation}\label{Eq:phiAE}
\ket{\psi}_{AE}=\frac{1}{2}(\ket{00}\ket{0}+\ket{01}\ket{1}+\ket{10}\ket{+}+\ket{11}\ket{-}),\nonumber
\end{equation}
and the prepared state $\rho_{AE}'$ can be obtained by measuring Alice's subsystem of $\ket{\psi}_{AE}$ in the computational basis, $I =\{\ket{00},\ket{01},\ket{10},\ket{11}\}$.
Note that, as we consider the case where Alice has two qubits in her system, the randomness against Eve should be $0\le R_I \le 2$. Therefore, Eve's information about Alice's measurement outcome is equivalent to consider the coherence of
\begin{equation}
\begin{aligned}
	&\rho_{A}\\
&=\frac{1}{4}((\ket{00}+\frac{1}{\sqrt{2}}(\ket{10}+\ket{11}))(\bra{00}+\frac{1}{\sqrt{2}}(\bra{10}+\bra{11}))\\
&+(\ket{01}+\frac{1}{\sqrt{2}}(\ket{10}+\ket{11}))(\bra{01}+\frac{1}{\sqrt{2}}(\bra{10}+\bra{11}))).
\end{aligned}
\end{equation}
The two quantum coherence measures are  $R_I^Q(\rho_A)=1$ and $R_I^C(\rho_A)=3/2$, being the latter obtained via numerical optimization. Hence, the quantum discord of the post-measurement state
 is $D_E(\ket{\psi}_{AE})=1/2$, measuring how much extra information Eve can obtain by performing coherent measurements than independent measurements of multiple copies of $\rho_{AE}'$.

\section{Conclusion}\label{concl}
Given the twofold uncertainty of a quantum measurement, we provided an operational interpretation to the genuinely intrinsic randomness about a measurement performed by an observer Alice, which we quantify with the relative entropy of coherence, as the minimum uncertainty about the outcome by a quantum correlated party Eve. We then compared the result to an alternative strategy to quantify quantum coherence by a convex roof extension of the Shannon entropy. The gap between the two strategies is equal to the discord of the bipartite state shared by Alice and Eve. The result provides a new link between single system quantumness and quantum correlations even in separable states, which was inspired by previous studies on the trade-off between local and global quantum properties\upcite{Girolami13,entco,Yao15,ma2015converting,Streltsov16,Chitambar16,Hu17relative,PhysRevX.7.011024}.
Following this line of thinking, other interesting scenarios where the interplay between coherence and correlations should be investigated is in the context of physical limits to privacy and to communication, e.g., data hiding protocols\upcite{DiVincenzo04,enigma,datta,molmer,boixo2011quantum}. Another potential avenue of further research is the extension of the result to the multipartite setting, i.e., to determine a link between local coherence and genuine multipartite quantum correlations. Note that in this paper, we only consider the uncertainty of a projective measurement (computational basis measurement), it is also interesting to explore the uncertainty of a general quantum measurements, instead of projective measurement\upcite{cao2015loss}. Quantum coherence is also connected to the generated randomness with the extraction process\upcite{zhao2019one,PhysRevA.97.012302}.

\section*{Acknowledgements}
This work was supported by the National Natural Science Foundation of China Grants No.~11875173 and No.~11674193, the National Key R\&D Program of China Grants No.~2017YFA0303900 and No.~2017YFA0304004, the Zhongguancun Haihua Institute for Frontier Information Technology, the EPSRC (Grant No. EP/L01405X/1), and the Wolfson College, University of Oxford.


\bibliographystyle{IEEEtran}
\bibliography{IEEEabrv,bibRQ}


\end{document}